\documentclass[aps,prl,reprint,floatfix,notitlepage,balancelastpage,twocolumn,showpacs]{revtex4}
\usepackage[dvips]{graphicx}
\usepackage{float,amsmath,amssymb,bm,color}
\date{\today}

\newcommand{\vf}{v_{\rm F}}

\newcommand{\br}{{\bf r}}

\newcommand{\be}{\begin{equation}}
\newcommand{\ee}{\end{equation}}
\newcommand{\bea}{\begin{eqnarray}}
\newcommand{\eea}{\end{eqnarray}}
\newcommand{\bse}{\begin{subequations}}
\newcommand{\ese}{\end{subequations}}

\begin{document}

\title{Inhomogeneous Superconducting States of Mesoscopic Thin-Walled Cylinders in External Magnetic Fields}

\author{K. Aoyama$^{1,2,3}$, R. Beaird$^2$, D. E. Sheehy$^2$, and I. Vekhter$^2$}

\affiliation{${}^1$ The Hakubi Center for Advanced Research, Kyoto University, Kyoto 606-8501, Japan\\
${}^2$ Department of Physics and Astronomy, Louisiana State University, Baton Rouge, LA 70803\\
${}^3$ Department of Physics, Kyoto University, Kyoto 606-8502, Japan
}

\begin{abstract}
We theoretically investigate the appearance of spatially modulated
superconducting states in mesoscopic superconducting
thin-wall cylinders in a magnetic field at low temperatures. Quantization of the
electron motion around the circumference of the cylinder leads to a discontinuous evolution of the
spatial modulation of the superconducting order parameter
along the transition line $T_c(H)$. We show that this  discontinuity leads to the non-monotonic behavior of the specific heat jump at the onset of superconductivity as a function of temperature and field. We argue that this geometry provides an excellent opportunity to directly and unambiguously detect distinctive signatures of the Fulde-Ferrell-Larkin-Ovchinnikov modulation of the superconducting order.
%Impurity effects on this quantum effect are also discussed.
\end{abstract}
\pacs{xxx}
\maketitle

%%%%%% introduction %%%%%%%%

Mesoscopic systems both serve as a platform to investigate fundamental quantum physics of solids and are a testing ground for potentially transformative future devices~\cite{QHE, qdot, topological, Majonara, LP}. Of special interest in this context are interacting systems exhibiting interplay of the collective emergent properties with the quantum physics of single particles. Small superconducting samples of nontrivial topology provide an example of such interplay since the global phase of the pair condensate and the phases of single particle wave functions respond differently to the applied magnetic field.

In mesoscopic superconducting rings of radius $R\sim\xi_0$, where
$\xi_0 \equiv \vf/2\pi T_{c}(H=0)$  is the superconducting coherence
length, and $v_F$ is the Fermi velocity, this leads to a doubling of
the period of the oscillations of the transition temperature, $T_c$,
as a function of the magnetic flux, $\Phi$, through the ring, relative
to the well-known Little-Parks (LP) effect~\cite{LP, ring, Goldbart}. The
small ring radius, $R$, implies that each single electron state can be
labeled by its angular momentum, $n$, in units of $\hbar$, and each
particle acquires an additional phase due to the magnetic flux,
$\Phi=\pi R^2 H$ when circling around the ring. In the absence of the
field the wave function of the Cooper pair has a net zero angular
momentum, as the time-reversed states with $n_1=-n_2$ form a bound
state. In contrast, under the applied field the non-zero quantum
number $l=n_1+n_2$ partially compensates the net flux and maximizes
the transition temperature, $T_c$. Therefore, for a small ring~\cite{Kulik,Goldbart}, $T_c$ is
a periodic function of $\phi=\Phi/(2\Phi_0)$ with the flux quantum
$\Phi_0=hc/2e$, while for a large ring the periodicity is solely due to
the flux experienced by an electron {\em pair} with charge $2e$,
i.e. $T_c$ is a periodic function of $\Phi/\Phi_0$ (LP effect).

Under these assumptions, there is no overall suppression of the maximal
$T_c$ as the magnetic field increases: at integer values of $\phi$ the
orbital coupling of the field to the  individual electrons can be
fully compensated by the finite angular momentum $l$ of the Cooper
pair. The destruction of superconductivity in such a geometry must
then occur via paramagnetic (Zeeman) coupling of the electron spins to
the field, which raises the energy of the singlet bound state:
inclusion of this coupling is essential for developing a complete
picture.  Consequently in this Letter we consider the combined effect
of the orbital and Zeeman effect on superconductivity, and analyze a
mesoscopic thin-walled cylinder with the field along the axis. The
 cylinder geometry
allows the formation of a spatially-modulated~\cite{FF,LO} (Fulde-Ferrell-Larkin-Ovchinnikov, FFLO)
superconducting state that enables pairing under high Zeeman field. We
show below that a) this state occurs even if the cylinder is made out of materials where superconductivity is not paramagnetically limited in the bulk; b) signatures of such a state are much more prominent and
easily identified in this geometry with $R\sim\xi_0$ than either in bulk
materials or flat thin films, and therefore mesoscopic systems offer a unique chance to
detect the FFLO state that has remained elusive for nearly 50 years
since it was first predicted.

\begin{figure}[t]
\begin{center}
\includegraphics[width=\columnwidth]{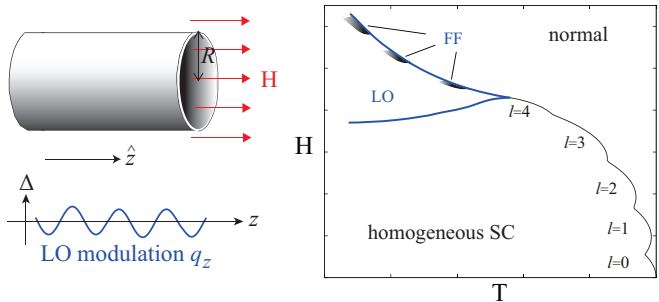}
\caption{(Color online) Superconducting thin-wall cylinder with a small radius $R$ in a magnetic field parallel to the cylinder axis ($z$-axis). Sketch shows the spatial modulation of the order parameter in the {\it LO} phase. Right-panel: phase diagram in the $H$-$T$ plane, showing regions of the normal phase, homogeneous
($q_z=0$) superconductor, and spatially-modulated LO phase, with
phase modulated FF states indicated by the shaded regions. \label{fig:pd}}
\end{center}
\vspace{-.5cm}
\end{figure}

Our main results are shown in Fig.~\ref{fig:pd}. While at low fields
the variation of the transition temperature $T_c(H)$ is well described
by the Little-Parks (LP) periodicity, at higher fields there is an extended
region in which the superconducting order parameter is
modulated along the cylinder axis.  Near the phase boundary $T_c(H)$ in this regime, the
superconducting phase exhibits alternating regions of phase-modulated
(FF) and amplitude-modulated (LO) order; however, the LO phase becomes stable
at lower  $T$.
The wavevector, $q_z$, of this modulation has non-analytic dependence
on $H$, due to the interplay between
the finite size effects and the LP oscillations.
The
heat capacity jump at $T_c(H)$ varies dramatically along this sequence of
transitions (in contrast to the smooth evolution at temperatures where the FFLO modulation is absent), enabling a direct identification of the modulated states.

Although the possibility of FFLO states has been discussed in bulk
materials such as the heavy-fermion superconductor CeCoIn${}_5$
\cite{Bianchi03,ultrasound,Kumagai_11} and organic superconductors
\cite{kappa-BEDT,lambda-BEDT,beta-ET, TMTSF}, it is difficult to design a ``smoking gun'' experiment that unequivocally points towards such a state. In real bulk systems both orbital and paramagnetic coupling suppress superconductivity, and inhomogeneity arises due to both. The former effect leads to proliferation of the vortices. Each vortex contains exactly one flux quantum for the Cooper pairs, $\Phi_0$, which corresponds to a 2$\pi$-phase winding of the superconducting order parameter around the vortex core. In that sense a thin-walled ring or cylinder can be viewed as a coreless ``supervortex'' of multiple flux quanta, with a phase winding $2\pi l$. In contrast, paramagnetic pairbreaking favors FFLO states. Recall that in a singlet superconductor in the absence of Zeeman splitting the Cooper pair comprises electrons in time-reversed states, which have equal energies,
and therefore are unstable towards formation of a bound state.
With paramagnetic coupling the states with opposite spins have equal energies if they have a field-dependent momentum mismatch $q$, and the modulation of the FFLO state originates from this finite center-of-mass momentum (CMM) of the Cooper pairs.
On a 1D ring the pair CMM is equivalent to the net phase winding, so that a different geometry is needed to distinguish the modulated states.

We consider a long hollow superconducting cylinder of radius $R$ and thickness $d\ll\xi_0$, which, in the absence of a magnetic field, is described
by the Hamiltonian
\begin{eqnarray}\label{eq:H_0}
{\cal H}&=&\sum_{\sigma, {\bm p}}\xi({\bm p}) {\hat c}^\dagger_{{\bf p},\sigma}
{\hat c}_{{\bf p},\sigma}
 -\lambda\sum_{\bf q} {\hat B}^\dagger({\bf q}) \, {\hat B}({\bf q}), \nonumber\\
{\hat B}({\bf q}) &=& \frac{1}{2}\sum_{{\bf p},\alpha,\beta}  (-i \, \sigma_y)_{\alpha, \beta} \, {\hat c}_{-{\bf p}+\frac{\bf q}{2},\alpha} {\hat c}_{{\bf p}+\frac{\bf q}{2},\beta}\,.
\end{eqnarray}
Here ${\hat c}_{{\bf p},\alpha}$ is the annihilation operator for an electron with momentum ${\bf p}$ and spin projection $\alpha$, $\lambda$ is the strength of the pairing interaction, and we assumed singlet $s$-wave superconductivity. For small $R$ the motion around the circumference of the cylinder is quantized, while the momentum along the axis ($z$) is continuous, so that ${\bf p}=(m/R, p_z)$ with $m$ integer, and the quasiparticle energy takes the form
\begin{equation}
\label{eq:dispersion}
\xi({\bf p}) = \frac{1}{2M}\Big[p_z^2+\Big( \frac{m}{R}\Big)^2 \Big]-\mu\,, %\quad (m:integer)\,,
\end{equation}
where $M$ is the electron mass and $\mu$ is a chemical potential. Here the summation over the momenta means $\sum_{\bf p}=\left[(2\pi)^2R\right]^{-1}\sum_{m \in Z} \int dp_z$.

The magnetic field threading the cylinder along its axis leads to a Zeeman splitting of the single particle energy levels by $H_Z=\sum_{\sigma, {\bm p}}\sigma \, h  \, {\hat c}^\dagger_{{\bf p},\sigma}
{\hat c}_{{\bf p},\sigma}$, where $h=\mu_{\rm B} \,g H$,  $\mu_{\rm B}$ is the Bohr magneton, and $g$ is the $g$-factor of the quasiparticles in the crystal. At the same time the momentum operator has to be replaced by its gauge-invariant counterpart, $\widehat{\bm p}\rightarrow \widehat{\bm p}+|e| {\bm A}$. For our model of a thin-walled cylinder the vector potential ${\bf A} = - \, {\hat \varphi} \, H \, R/2 $, and $\varphi$ is the azimuthal angle around the cylinder. Hence $|{\bf A}|=const$ on the cylinder.

In the superconducting (SC) phase the pair field $\Delta({\bf q})=\lambda \langle {\hat B}({\bf q}) \rangle$, where
$\langle \ldots \rangle$ denotes the thermal average, acquires a non-zero value.
Due to the cylindrical geometry, $\Delta(\br)$ can be expanded in the Fourier series, %transforms,
\begin{equation}\label{eq:SCgap_cylinder}
\Delta({\bf r})=|\Delta_0|\sum_{q_z}\sum_{l\in Z} C_{l,q_z} e^{i \, l\varphi} e^{i \, q_z \, z}.
\end{equation}
The uniform SC state at $H=0$ only has $C_{0,0}\neq 0$, while a single
component $C_{l,0}$ with flux-dependent $l \neq 0$ characterizes the
SC transition under orbital coupling to the field and gives rise to
the Little-Parks effect. If the cylinder were to unfold into a
two-dimensional (2D) plane, $l$ and $q_z$ would become components of a
2D vector $\bm q$, and in response to a Zeeman field a state with $\bm
q\neq 0$ would be realized. The cylindrical geometry is
unique since $l\neq 0$ gives the magnetic flux through the cylinder,
%under both orbital and Zeeman coupling,
and therefore the %additional
FFLO modulation is only along the axis, $q_z\neq 0$. Near the transition the linearized gap equations for different $l$, $|q_z|$ decouple~\cite{Matsuda2007,Loder2012}, and hence superconducting states appear either with a single $C_{l, \bm q_z}$ (phase modulated, $\Delta({\bf r})=\Delta_0 e^{i l\varphi} e^{i\bm q_z z}$, FF), or with $C_{l, q_z}=C_{l, -q_z}$ (amplitude modulated, $\Delta({\bf r})=\Delta_0 e^{i l\varphi} \cos(\bm q_z z)$, LO, see Fig.~1).  Thus, this
setting has an advantage over other ways to achieve FFLO states (such as
bulk paramagnetically-limited superconductors, imbalanced fermionic atomic gases~\cite{RS2010} or thin SC films), in which
modulation direction is arbitrary, and therefore complex states
may be favored~\cite{Shimahara, Combescot, Houzet-Buzdin, Adachi, Ikeda, LeoPRA}.

To study the transition between the normal and  SC states we use the
Ginzburg-Landau (GL) expansion of the free energy, ${\cal F}_{\rm
  GL}=a^{(2)}(l,q_z, T, H) |\Delta_0|^2 + a^{(4)}(l,q_z, T,
H)|\Delta_0|^4$, At each $H, l, q_z$, the temperature $T$  where $a^{(2)}(l,q_z, T, H)$ becomes negative (if it
exists) indicates a
putative second order transition from the normal into the SC state
with given values of $l, q_z$. The highest of these temperatures is
the physical transition point $T_c(H)$ into a state with the
corresponding $l, q_z$. The necessary condition for the continuous
transition is that the quartic coefficient $a^{(4)}(l,q_z, T_c(H), H)$
remains positive at the transition point.

We evaluate the coefficients $a^{(2)}$ and $a^{(4)}$ using the Green's function formalism for the Hamiltonian,
Eq.~\eqref{eq:H_0} with the Zeeman and orbital coupling terms. 
We note here that a similar setup was considered in Ref.~\onlinecite{Zyuzin}
using a phenomenological GL expansion that is valid only in the long-wavelength
modulation limit.  Due to the neglect
of the field dependence of the coefficients of the GL expansion, lack of connection with a microscopic model Hamiltonian, and the
assumption of a small modulation wave vector, that approach failed to
capture any of the physics found in this Letter, and led the authors of
Ref.~\onlinecite{Zyuzin} to focus on the fluctuation contribution to
the specific heat as the main observable. Our analysis below shows that the
``mean field'' features of the transition, when analyzed properly, strongly reflect the
interplay of the quantization of single electron motion and spatial
modulation of the SC order. We use a quasi-classical approximation for the normal state Green's function  $-\langle T_\tau {\hat \psi}_\sigma({\bf r},\tau) \,
{\hat \psi}^\dagger_{\sigma} ({\bf r}',0) \rangle \simeq
T\sum_{\varepsilon_n} \,e^{-\varepsilon_n \tau} \, {\cal
G}_{\varepsilon_n, \sigma}({\bf r}-{\bf r}') \, e^{i |e| \int_{{\bf
r}}^{{\bf r}'} d{\bf s}\cdot {\bf A}({\bf s})}$, where the integral in the exponent is evaluated along a straight line. This approximation smears out the even-odd flux periodicity for a 1D ring, but this periodicity is already broken by Zeeman coupling, and hence the approximation is adequate for our goals.   We obtain
 for the quadratic term
\begin{eqnarray}\label{eq:GL}
&&a^{(2)}|\Delta_0|^2 = \int d{\bf r}\, \Delta^\ast({\bf r})\Big( \frac{1}{\lambda}-\frac{T}{2}\sum_{\varepsilon_n,\sigma} \sum_{\bf p} \hat{K}(\varepsilon_n,\sigma) \Big)\Delta({\bf r}), \nonumber\\
&&\hat{K}(\varepsilon_n,\sigma) = {\cal G}_{\varepsilon_n, \sigma}({\bf p}) \, {\cal G}_{-\varepsilon_n, -\sigma}(-{\bf p}+{\bf \Pi}),
\end{eqnarray}
where ${\cal G}_{\varepsilon_n, \sigma}({\bf p})=(i\varepsilon_n - \xi({\bf p})+\sigma h)^{-1}$ is the Fourier transform of ${\cal G}_{\varepsilon_n,\sigma}({\bf r})$, $\varepsilon_n$ is a fermionic Matsubara frequency, and ${\bf \Pi}=-i\, \nabla +2|e|{\bf A}$.
The full expression for the quartic term is given in the supplementary information~\cite{suppl}.

The effects due to small ring size $R\sim\xi_0$  are contained in the discrete sum over integers $m$ in $\sum_{\bf p}$.  We use the Poisson summation formula~\cite{Goldbart}, $\sum_{m \in Z} \delta(x-m)=\sum_{k \in Z} e^{i \, 2\pi k\,x}$ to elucidate these effects:
$k=0$ gives the continuum result for a 2D superconductor, and higher order terms, $k\geq 1$, account for the finite size corrections. After a straightforward  calculation, we find

%%%%%%%%%%%%%%%%%%%%%%%%%%%%%%%%
\begin{widetext}
\begin{eqnarray}\label{eq:quadratic}
%&&\frac{a^{(2)}}{M/2\pi}
&&a^{(2)}(l,q_z,T, H)\simeq \frac{M}{2\pi}%\sum_{l,q_z}  |C_{l,q_z}|^2
\Bigg[ \ln \Big(\frac{T}{T_c}\Big) + \psi\Big(\frac{1}{2} \Big) - \frac{1}{4} \sum_{s_\varepsilon,\sigma=\pm 1} \int_0^{2\pi} \frac{d\varphi_p}{2\pi} \psi\Big(\frac{1}{2} -\frac{i \, s_\varepsilon}{4\pi T}\big[2\sigma h - {\bf v}_F\cdot {\bf Q} \big] \Big) \\
&+&2 \, \sum_{k >0}\int_0^{2\pi} \frac{d\varphi_p}{2\pi}  \cos\Big(2\pi \, k \, R \,p_F \cos(\varphi_p) \Big) \sum_{n>0}\Big(\frac{ e^{-\frac{R}{\xi_0} \pi(2n+1) k |\cos(\varphi_p)|} }{n+\frac{1}{2}} - \frac{1}{4}\sum_{s_\varepsilon,\sigma}\frac{e^{-\frac{R}{\xi_0} \frac{T}{T_c}\pi  (2n+1) k |\cos(\varphi_p)|} }{n+\frac{1}{2}-\frac{i \, s_\varepsilon}{4\pi T}[2 \sigma h - {\bf v}_F\cdot {\bf Q}]}\Big)\Bigg],  \nonumber
\end{eqnarray}
\end{widetext}
where we defined the product
%%%%%%%%%%%%%%%%%%
\begin{equation}
{\bf v}_F \!\cdot\! {\bf Q} = 2\pi T_c \Big( \xi_0 q_z \sin(\varphi_p)\!+\!\Big[l\!-\!\frac{\Phi}{\Phi_0} \Big]\frac{\xi_0}{R} \cos(\varphi_p) \Big).
\label{eq:vq}
\end{equation}
%%%%%%%%%%%%%%%%%%
\begin{figure}[b]
\begin{center}
\includegraphics[scale=0.62]{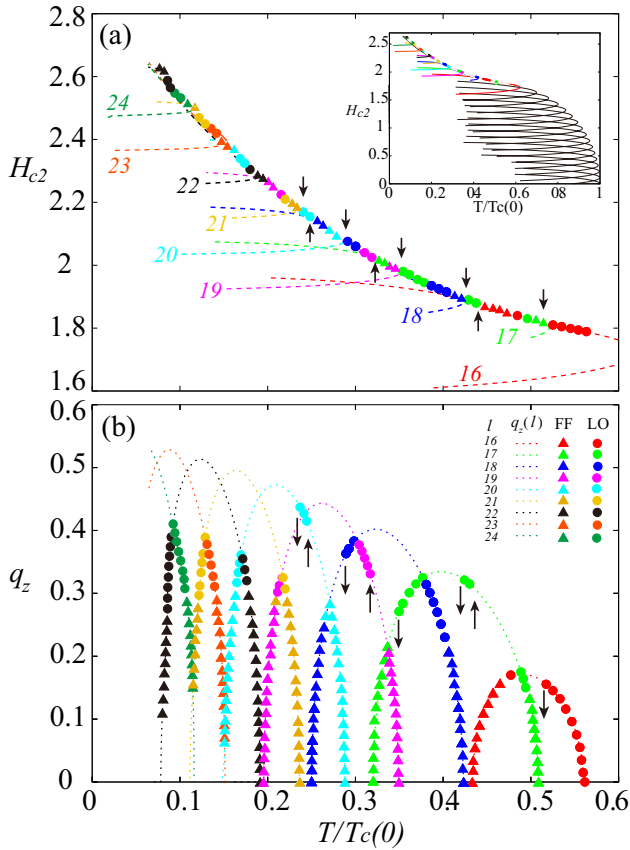}
\caption{(Color online) Structure of the modulated state for $R/\xi_0=3$ and $\alpha=0.6$.
(a) The upper critical fields $H_{c2}$ normalized by $\Phi_0/\pi\xi_0^2$. Inset and dashed lines show $H_{c2}(T)$ for a given angular momentum $l$ pairing state as indicated. Solid symbols denote the physical transition. Note the switching between the FF (triangles) and LO (circles) states along the transition line.  (b) FFLO modulation wave vector $q_z$ normalized by $1/\xi_0$ for each $l$ (dotted lines), and at the physical transition as in panel (a). The circles (triangles) denote the stability regions of the LO (FF) state. Note the non-analytic behavior of $q_z$ exhibiting kinks and discontinuous jumps at temperatures denoted by arrows.
\label{fig:R3005-3Hc2-q}}
\end{center}
\end{figure}
Here,  $\psi(z)$ is the digamma function and  the Fermi momentum
is $p_F=\sqrt{2M \mu}$.
 In the second line of Eq.~(\ref{eq:quadratic}) we neglected terms of order $T_c^2/\mu^2$.
Because of the exponential decay of the terms with increasing $k>0$ in the last term of  Eq.~(\ref{eq:quadratic}), below we keep only the
first finite size correction, $k=1$.
We checked that incorporating higher $k$ does not qualitatively change our results.
%\blue{Although cylindrical geometries have been discussed elsewhere \cite{Zyuzin}, there is no theoretical study taking account of both this small size effect and the Zeeman field in its microscopic model.}

Fig.~\ref{fig:R3005-3Hc2-q} shows the upper critical field and the parameters $l,q_z$ of the modulation of the superconducting order parameter at transition. Hereafter we consider $\xi_0 p_F=100$, and $R\simeq 3 \xi_0$. In the cylindrical geometry the inhomogeneous superconducting states emerge even for the materials that do not support FFLO modulation in the bulk:  we present the results for the paramagnetic parameter $\alpha_M=g \mu_B
\Phi_0/(\pi \xi_0^2 \, T_c)=0.6$, which corresponds to the Pauli limiting field $H_P\approx 4 H_{c2}^{\rm{orb}}$, so that the bulk material is a conventional orbital-limited type-II superconductor. The scalloped shape of the boundary of the superconducting region is the consequence of the LP effect, and the overall suppression of $T_c$ with increased field is due to the paramagnetic pairbreaking. Below a characteristic temperature, which is non-universal and different from the $T^\star=0.56T_{c0}$ for bulk Pauli limited superconductors, the inhomogeneous pairing along the cylinder axis ($q_z\neq 0$) becomes advantageous, and the FFLO state appears.

Fig.~\ref{fig:R3005-3Hc2-q} shows that the modulation wave vector,
$q_z$, along the transition line exhibits a ``sawtooth'' pattern,
quite distinct from the uniform increase in $q$ in the standard
picture of Pauli-limited superconductors.  This feature is due to the
effective discretization of the modulation in Eq.~\eqref{eq:vq}. For a
2D sheet the role of the winding number $l$ is taken by a continuous
variable $q_x$, and it is the net $q=\sqrt{q_z^2+q_x^2}$ that ensures
matching of the energy of the two electrons in a Cooper pair with the
center of mass momentum $q$. In contrast, in the cylindrical geometry
the choice of $l$ is determined by the net flux, $\Phi$, and therefore
the momentum $q_z$ adjusts to this selection, and exhibits
discontinuities at the points where transitions between winding
numbers $l$ and $l+1$ occur. The detailed balance between $q_z$ and
$l$ depends on the finite size quantum correction term, second line in
Eq.~(\ref{eq:quadratic}). Note that the prefactor of the first
non-vanishing $k$ term, $\cos[2\pi R \, p_F \cos(\varphi_p) ]$, has the
same angle-dependence in the momentum space as the the LP term,
$|l-\Phi/\Phi_0|\cos(\varphi_p)/R$, and is out of phase with the FFLO
modulation that enters with $\sin (\varphi_p)$ in
Eq.~\eqref{eq:vq}. Consequently the details of the switching between
values of $l$ and $q_z$ depend on the value of $Rp_F$ (close to integer
vs. half-integer), but the qualitative picture remains unchanged.

\begin{figure}[t]
\begin{center}
\includegraphics[scale=0.64]{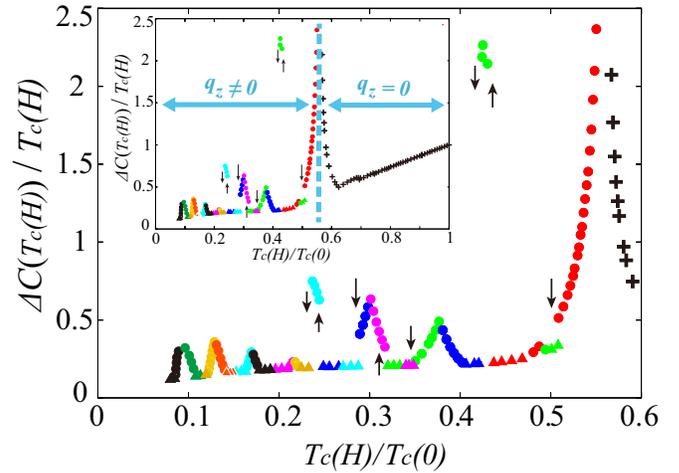}
\caption{(Color online) The specific heat jump at the onset of the modulated SC order in field, $\Delta C(T_c(H))/T_c(H)$. % along the $H_{c2}(T)$ curve, where the same parameters and colored-symbols as
The notations are the same as in Fig.\ref{fig:R3005-3Hc2-q}. Inset: the same over the entire temperature range of the superconducting transition. The non-monotonous behavior appears {\em only} for the FFLO state, $q_z\neq 0$.
%shows the evolution of the specific heat near the discontinuity in the modulation $q_z$ at $T/T_c=0.29$. %$C_s$ and $\Delta C/T$
%All quantities are normalized by their values at $T_c(H=0)$.
\label{fig:R3state}}
\end{center}
\end{figure}

The discontinuous behavior of the modulation $q_z(T)$
is reflected in the experimental properties that allow unambiguous determination of the modulated state.
Fig.~\ref{fig:R3state} shows the specific heat jump at the superconducting transition for different fields. We verified that the quartic term, $a^{(4)}(l,q_z, T_c(H), H)$,  remains positive along the entire transition line, and therefore the transition is always of second order.
The favored state is determined by comparing the magnitude of the quartic term for the FF and the LO states: A smaller value corresponds to the greater condensation energy and a more stable phase. We find that in the vicinity of the discontinuous drop of the modulation $q_z$ the FF state is favored, and is superseded by the LO state as $q_z$ increases within the realm of each fixed winding number $l$. The specific heat jump at the transition is given by (we omit full labels for brevity) $\Delta C /T_c(H) = ([a^{(2)}]^\prime)^2/2 a^{(4)}$ evaluated at $T_c(H)$, where $[a^{(2)}]^\prime=(\partial a^{(2)}/\partial T)$.%_{T=T_c(H)}$.
The heat capacity exhibits significant enhancement on transitions between different winding numbers. It is important to note that this non-monotonous behavior of the specific heat jump only appears when the transition is into FFLO state, at low temperatures and high fields. At higher $T$, when the transition is into the superconducting state with $q_z=0$, the specific heat jump at the transition varies smoothly, see the inset of Fig.~3. The enhancement of the specific heat jump in the hollow cylinder geometry can be detected, for example, by the ac calorimetry technique, and therefore can serve as experimental proof of the existence of the FFLO-like modulations of the superconducting order in mesoscopic cylinders.

It is likely that in the experimental
realization of the  proposed geometry the superconductor will be
disordered. We checked that the modulated states, and the non-monotonic behavior of $q_z(T)$
are robust against moderate impurity scattering~\cite{suppl}
The former result is consistent with the conclusions of
Refs.~\onlinecite{impurity_Agterberg,impurity_Vorontsov}. FFLO modulation disappears at strong disorder when transport becomes diffusive~\cite{impurity}.

To conclude, we find novel spatially-inhomogeneous superconducting
states, exhibiting both the Little-Parks and FFLO phenomenology, can emerge due to the
vector potential and Zeeman coupling induced by a magnetic field threading
a thin hollow cylinder.  Our principal motivation was conventional superconductors, for which the coherence
length can be well in excess of 1000\AA.  In this setting, the relevant sample sizes are
experimentally accessible and we believe that the predicted
variations in the specific heat jump can be found under realistic
conditions, providing a possible \lq\lq smoking-gun\rq\rq\ experiment for detecting
the FFLO state.
In principle,
cold atomic gases in a cylindrical geometry and coupled to a light-induced
artificial magnetic field could realize a similar phase diagram
and FFLO state
~\cite{LinPRL2009}.  However, given the difficulty of directly measuring the heat-capacity jump in a trapped cold atomic
gas, it may currently be easier to detect the effects in small-size superconducting systems.

This work is supported by NSF via Grant No. DMR-1105339 (K. A. and
I. V.) and  Grant No. DMR-1151717 (D.E.S.). Portions of this research
were conducted with high performance computing resources provided by
the Center for Computation and Technology at LSU and Louisiana Optical
Network Initiative.

\begin{widetext}
\section{Supplemental Material for "Inhomogeneous Superconducting States of Mesoscopic Thin-Walled Cylinders in External Magnetic Fields"}

\subsection{1. Quartic Term in the Ginzburg-Landau Expansion}
The quartic term in the expansion of the free energy is given by a combination of four single-particle propagators
\begin{eqnarray}
a^{(4)}(l,q_z,T,H)|\Delta_0|^4&=&\frac{1}{2}\int d{\bf r} \, \hat{K}_4({\bf \Pi}_i) \Delta^\ast({\bf s}_1)\Delta({\bf s}_2) \Delta^\ast({\bf s}_3)\Delta({\bf s}_4) \big|_{{\bf s}_i\rightarrow{\bf r}},
\nonumber\\
\hat{K}_4({\bf \Pi}_i)&=& \frac{T}{2}\sum_{\varepsilon_n,\sigma} \sum_{\bf p} \, {\cal G}_{\varepsilon_n, \sigma}({\bf p}) \, {\cal G}_{-\varepsilon_n, -\sigma}(-{\bf p}+{\bf \Pi}^\dagger_1)
%\nonumber\\&\times&
{\cal G}_{-\varepsilon_n, -\sigma}(-{\bf p}+{\bf \Pi}_2) \, {\cal G}_{\varepsilon_n, \sigma}({\bf p}+{\bf \Pi}^\dagger_3-{\bf \Pi}_2),
\end{eqnarray}
where ${\bf \Pi}_i=-i\nabla_{\bm s_i}+ 2|e|{\bm A}$ acts on $\Delta({\bf s}_i)$ \cite{GL}.
The summation over ${\bf p}$ is performed in the same manner as that used in obtaining Eq.~(5) in the text, and we find
\begin{eqnarray}\label{eq:quartic}
&&a^{(4)}(l,q_z,T,H) = \frac{1}{2}\frac{M}{2\pi}\sum_{q_1+q_3=q_2+q_4} C_{l,q_1}^\ast C_{l,q_2} C_{l,q_3}^\ast C_{l,q_4}  %\nonumber\\
%&&\times
\sum_{k \in Z} \int_0^{2\pi} \frac{d\,\varphi_p}{2\pi} \, \exp\Big[i \, 2\pi k R \, p_F \cos(\varphi_p) \Big]
\nonumber\\
&&\times \prod_{j=1}^3\int_0^{\infty} d\rho_j \frac{2 \pi T \, \cos\big( 2 h \, [\sum_{i=1}^3\rho_i ] \big)}{\sinh\big[2\pi T ([\sum_{i=1}^3\rho_i]+\frac{1}{2}\frac{R}{\xi_0}|k \cos(\varphi_p)|)\big]}
%\nonumber
\label{eq:quartic}
\\ \nonumber
&& \times \Big[ \cos\big( {\bf v}_F\cdot {\bf Q}_1 (\rho_1 + \rho_2) - {\bf v}_F\cdot {\bf Q}_2 \rho_2 + {\bf v}_F\cdot {\bf Q}_3(\rho_2+\rho_3) \big)
%\nonumber\\
%&&
+\cos\big( {\bf v}_F\cdot {\bf Q}_1 \rho_1 + {\bf v}_F\cdot {\bf Q}_2 \rho_2 + {\bf v}_F\cdot {\bf Q}_3 \rho_3 \big) \Big],
\end{eqnarray}
where ${\bf Q}_i=\big(\frac{1}{R}[l-\Phi/\Phi_0], q_i \big)$, the Fermi velocity is ${\bf v}_F=2\pi T_c\xi_0 \big(\cos(\varphi_p),\sin(\varphi_p)\big)$, and we used the identity $\alpha^{-1}=\int_0^\infty d\rho \, \exp[-\alpha \, \rho] \, ({\rm Re}\alpha >0)$ to exponentiate the operators. We evaluate Eq.~(\ref{eq:quartic}) using the definitions of the FF and the LO states, wherein
\begin{eqnarray}
&&C_{l,q_i}=\delta_{q_i, q_z} \quad \big( \Delta({\bf r})=|\Delta_0|e^{i \, l\varphi} e^{i \, q_z \, z}\big)\,, \\ %\nonumber\\
&&C_{l,q_i}=\frac{\delta_{q_i, q_z} + \delta_{q_i, -q_z}}{\sqrt{2}} \quad \big(\Delta({\bf r})=|\Delta_0|e^{i \, l\varphi}{\sqrt 2}\cos( q_z \, z)\big)\,, 
\end{eqnarray}
are used for the FF and LO states, respectively.
%\end{widetext}

\subsection{2. Impurity Effects on the FFLO State}
\begin{figure}[t]
\begin{center}
\includegraphics[scale=0.45]{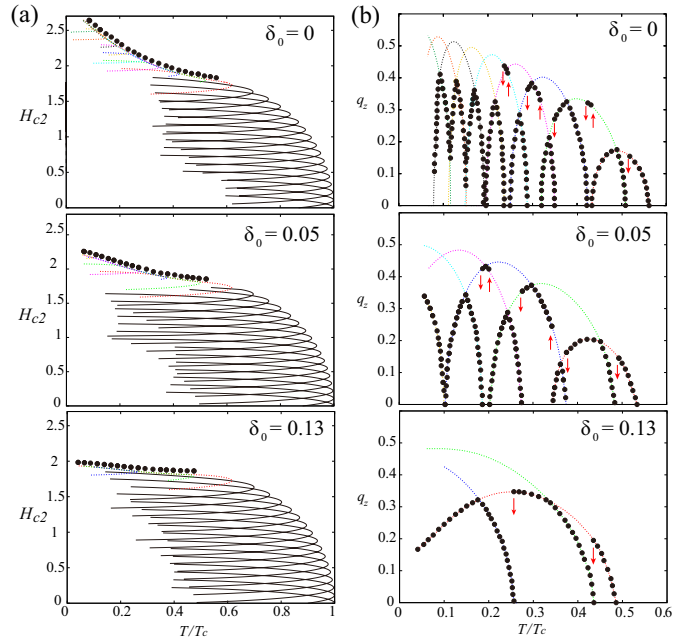}
\caption{Impurity effect on the modulated state for $R/\xi_0=3$ and $\alpha_M=0.6$. (a) The upper critical field $H_{c2}$, and (b) the FFLO modulation wave vector $q_z$ for $\delta_0=0$ (top), $\delta_0=0.05$ (middle), and $\delta_0=0.13$ (bottom). Black dots trace the FFLO stability region (a) and the corresponding physical modulation (b). The notations and normalizations are the same as in Fig. 2 in the body. \label{fig:fig_4}}
\end{center}
\end{figure}

It is well known that in a strongly disordered regime, when the single-particle propagation is diffusive, the FFLO state does not survive~\cite{impurity}. However, in many realistic situations, even though disorder leads to a finite lifetime of quasiparticles, the transport remain ballistic. We focus on this regime and explore how robust our conclusions are against a finite concentration of non-magnetic impurities. Those are described by an additional term in the Hamiltonian
\begin{equation}
%{\cal H}_{\rm imp} = \sum_\sigma\sum_{{\bf p},{\bf p}'} \, u({\bf p}-{\bf p}') \, {\hat c}^\dagger_{{\bf p},\sigma}{\hat c}_{{\bf p^\prime}},\sigma \,,
{\cal H}_{\rm imp} = \sum_\sigma\sum_{{\bf p},{\bf p}^\prime} \, V({\bf p}-{\bf p}') \, {\hat c}^\dagger_{{\bf p},\sigma}{\hat c}_{{\bf p^\prime},\sigma}\,,
\end{equation}
where $V({\bf q})$ is the Fourier transform of the scattering potential of a random ensemble of impurities, $V(\bm r)=\sum_i u(\bm r-\bm R_i)$, located at positions $\bm R_i$ with a net concentration $n_{imp}$. We further assume that individual impurity potential is short-ranged and isotropic, so that the $s$-wave scattering amplitude $u_0$ is dominant, $u({\bf q})\simeq u_0$, and assume a small phase shift of scattering (Born limit).
Within the Born approximation, computing the quadratic term in the Ginzburg-Landau expansion requires inclusion of the vertex corrections, and the result is obtained by replacing ${\hat K}(\varepsilon_n,\sigma)$ with ${\hat K}({\tilde \varepsilon}_n,\sigma)/[1-|u_0|^2 {\hat K}({\tilde \varepsilon}_n,\sigma)]$ in Eq.~(4) in the main text in analogy to Ref.~\cite{impurity_Agterberg}, where the renormalized frequency ${\tilde \varepsilon}_n = \varepsilon_n + {\rm sgn}(\varepsilon_n) /{2\tau}$, and the lifetime $\tau$ is  %\cite{Goldbart}
\begin{eqnarray}
\frac{1}{2\tau} &=& 2\pi T_c \, \delta_0 \Big(1 + 2 \, \sum_{k >0} \int_0^{2\pi} \frac{d\varphi_p}{2\pi}  \cos\big( 2\pi R \, p_F \, k \cos(\varphi_p)\big) %\nonumber\\
%&\times&
\exp \Big[-2\pi \frac{T}{T_c}  \frac{R}{\xi_0}\Big|\Big(n+\frac{1}{2} + \frac{\delta_0 T_c}{T}  \Big)k \cos(\varphi_p)\Big|\Big] \, \Big) .
\end{eqnarray}
The dimensionless parameter $\delta_0 =n_{imp} M \, |u_0|^2/(4\pi T_c)$ measures the strength of the impurity scattering.

Figure \ref{fig:fig_4} shows the upper critical field, $H_{c2}(T)$,  and the corresponding evolution of the modulation $q_z(T)$ for $\delta_0=0$ (clean limit), $\delta_0=0.05$ and $\delta_0=0.13$. At zero field, $T_c$ is not suppressed by nonmagnetic impurities as a manifestation of the Anderson's theorem~\cite{Anderson}. While the onset of the FFLO modulation is suppressed to lower temperature by impurity scattering, the ``sawtooth'' behavior in the $q_z(T)$ curve survives. This indicates that even in a moderately disordered superconductor, the qualitative features of our main conclusions persist. In particular, the non-monotonous evolution of the specific heat jump at the transition can still be observed in experiments, and serve as a strong evidence for the spatially inhomogeneous FFLO state. Note that, while we did not explicitly check at what impurity concentration the transition may become first order order, in known cases the nature of the FFLO-normal transition is not altered by Born impurity scattering~\cite{impurity_Agterberg, impurity_Vorontsov}. The situation may be different in the strong scattering limit, but we leave this subject for future studies. Our analysis here established that the conclusions in the main text of the paper are robust against moderate impurity scattering.

\end{widetext}

\end{document}